\documentclass[aps,twocolumn,prl]{revtex4-2}

\usepackage{xcolor,hyperref}
\hypersetup{
   colorlinks,
   linkcolor={blue!50!black},
   citecolor={blue!50!black},
   urlcolor={blue!80!black}
}

\usepackage{epsfig, amsmath}
\usepackage{bm}
\usepackage{soul}
\usepackage{hyperref}
\usepackage{footmisc}
\DeclareMathAlphabet{\mathitbf}{OML}{cmm}{b}{it}

\begin{document}

\title{Finite-size study of the athermal quasistatic yielding transition in structural glasses}

\author{David Richard}
\affiliation{Institute for Theoretical Physics, University of Amsterdam, Science Park 904, 1098 XH Amsterdam, The Netherlands}
\author{Corrado Rainone}
\affiliation{Institute for Theoretical Physics, University of Amsterdam, Science Park 904, 1098 XH Amsterdam, The Netherlands}
\author{Edan Lerner}
\affiliation{Institute for Theoretical Physics, University of Amsterdam, Science Park 904, 1098 XH Amsterdam, The Netherlands}

\maketitle

Upon external mechanical loading, amorphous solids such as foams, colloids, or molecular glasses exhibit a yielding transition from an elastic response at small deformations, to a fluidlike, predominantly plastic response at large deformations. The nature of this transition can vary across materials, and generally depends on glasses' formation history, in addition to external parameters such as the temperature $T$ and loading rate $\dot{\gamma}$~\cite{Schuh_review_2007,eran_prapplied_2016}.

\begin{figure}[ht!]
  \includegraphics[scale=0.9]{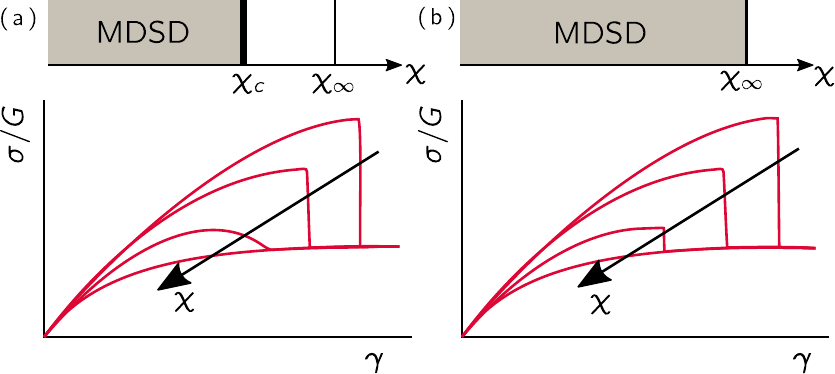}
  \caption{(a) The scenario proposed in Refs.~\cite{popovic2018elastoplastic,ozawa2018random}: continuous and discontinuous yielding are separated by a critical disorder $\chi_c\!<\!\chi_\infty$ represented by a finite stress overshoot. (b) The scenario proposed in Ref.~\cite{barlow2020ductile}: any vanishingly small stress overshoot is accompanied by a MDSD. The arrows indicate increasing structural disorder $\chi$.}
  \label{fig:intro}
\end{figure}

One well-defined limit of yielding popularized by computer simulations is the athermal ($T\!\to\!0)$, quasistatic ($\dot{\gamma}\!\to\!0$) limit~\cite{Malandro_Lacks,maloney2006amorphous}. In this limit, depending on glass-formation protocol, a \emph{macroscopic}, discontinuous stress drop (MDSD) can occur, which persists in the thermodynamic limit. The occurrence of a MDSD depends on the degree of mechanical disorder --- denoted by $\chi$ in what follows --- featured by as-cast glasses. Conflicting propositions were put forward in previous work regarding the \emph{minimal} $\chi$ required in order to observe a MDSD in sheared glasses: Refs.~\cite{ozawa2018random,popovic2018elastoplastic,ozawa2021rare} argue that there should exist a critical $\chi_c\!<\!\chi_\infty$ ($\chi_\infty$ is the maximal disorder that a glass can attain) above which no MDSD is observed, as illustrated in Fig.~\ref{fig:intro}a. On the other hand, Ref.~\cite{barlow2020ductile} provides evidence that MDSDs occur for any $\chi\!<\!\chi_\infty$ of as-cast glasses, see Fig.~\ref{fig:intro}b.

\begin{figure*}[t!]
  \includegraphics[scale=0.9]{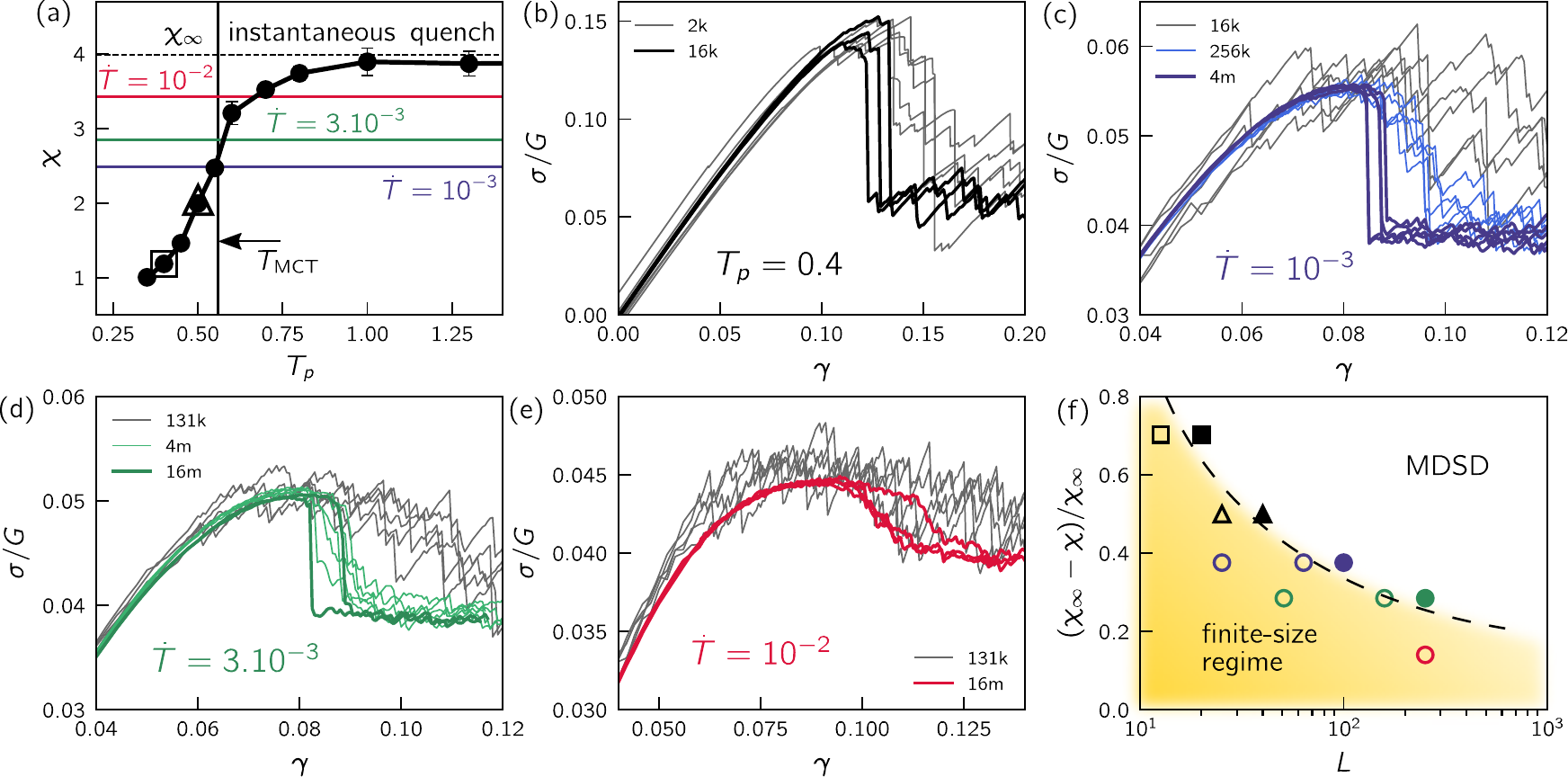}
  \caption{(a) As-cast disorder $\chi$ quantified via shear modulus fluctuations of our polydisperse glasses, plotted \emph{vs}.~preparation temperature $T_{\rm p}$. Colored horizontal lines indicate the degree of disorder featured by our finite $\dot{T}$ binary glasses, while the black dashed line corresponds to the maximal $\chi_\infty$ obtained with $\dot{T}\!\to\!\infty$ from a high-temperature melt. The vertical line indicates the mode coupling temperature $T_{\rm MCT}\!\approx\!0.56$. (b-e) Finite size study of the yielding transition for polydisperse glasses prepared at $T_p=0.40$ (b) and binary glasses prepared with $\dot{T}=10^{-3}$ (c), $3.10^{-3}$ (d), and $10^{-2}$ (e). (f) Phase diagram of the continuous/discontinuous yielding transition on the box length $L$ and relative disorder $(\chi_\infty\!-\!\chi)/\chi_\infty$ plane. Empty and filled symbols indicate continuous (finite-size regime) and discontinuous (MDSD) yielding,~respectively.}
  \label{fig:main}
\end{figure*}

In this Note we test the veracity of these two scenarios using the massively parallel LAMMPS library~\cite{plimpton1995fast}. We perform athermal quasistatic shear (AQS) of two models in three-dimensions (3D): a binary~\cite{lerner2018characteristic} and a polydisperse~\cite{lerner2019mechanical} glass. The binary glasses were cooled down to zero temperature using different quench rates $\dot{T}$, while our polydisperse systems were quenched from liquid configurations equilibrated at low temperatures $T_{\rm p}$ via SWAP Monte-Carlo~\cite{ninarello2017models}. In order to quantify the degree of mechanical disorder of our different models on the same footing, we employ Schirmacher's $N$-independent disorder quantifier $\chi\!\equiv\!\sqrt{N}\mbox{std($G$)/mean($G$)}$~\cite{Schirmacher_prl_2007} where $G$ is the shear modulus, $N$ is the system size, and we consider ensemble statistics~\cite{kapteijns2021elastic}. In Fig.~\ref{fig:main}(a), we show the as-cast disorder $\chi$ measured in our polydisperse glasses at various $T_{\rm p}$. We see that the polydisperse glasses' $\chi$ plateaus at high $T_{\rm p}$ and matches the disorder characterizing instantaneous quenches ($\dot{T}\!\to\!\infty$) of the binary glass. Furthermore, glasses prepared with our lowest quench-rate ($\dot{T}\!=\!10^{-3}$) show a similar disorder as polydisperse glasses quenched from a liquid equilibrated close to the mode coupling temperature $T_{\rm MCT}$.

Firstly, we shear low-$T_{\rm p}$ polydisperse glasses, see Fig.~\ref{fig:main}b. Consistent with Ref.~\cite{ozawa2018random}, we find that deeply annealed glasses ($T_{\rm p}\!=\!0.40$, empty square in Fig.~\ref{fig:main}a) exhibit a MDSD for systems of merely $N\!=\!16$k particles. Nevertheless, we do observe that smaller systems ($N\!=\!2$k) made with the same protocol ($T_{\rm p}\!=\!0.40$) do not feature a MDSD. We next shear our binary glasses that feature higher $\chi$ (see horizontal lines in Fig.~\ref{fig:main}a) and conversely show a much smaller stress overshoot. We clearly observe for both $\dot{T}\!=\!10^{-3}$ (Fig.~\ref{fig:main}c) and $\dot{T}\!=\!3.10^{-3}$ (Fig.~\ref{fig:main}d) a change from continuous to discontinuous yielding as the system size is increased. For the highest quench rate $\dot{T}=10^{-2}$ (Fig.~\ref{fig:main}e), the system size at which one would observe discontinuous yielding is expected (see Fig.~\ref{fig:main}f) to lie beyond the scope of atomistic simulations, but we can still appreciate its premise by comparing the stress-strain signals of $N\!=\!16$m with those of $N\!=\!131$k. Finally, we draw in Fig.~\ref{fig:main}(f) a soft boundary between continuous and discontinuous yielding transition in the box length $L$ and relative distance to maximal disorder $(\chi_\infty\!-\!\chi)/\chi_\infty$ plane. We see that the crossover $L^*(\chi)$ needed to observe MDSD grows exponentially as $\chi\!\to\!\chi_\infty$. This phase diagram is consistent with Ref.~\cite{shekhawat2013damage}. We very roughly estimate the minimal system size required to observe MDSD for $\dot{T}=10^{-2}$ at around $10^9$ particles.

Our results demonstrate that the athermal, yielding transition in the limit of vanishing strain rate exhibits significant finite-size effects as the degree of disorder approaches the maximal value $\chi_\infty$ obtained from an instantaneous quench of a high-temperature melt. On the basis of our data we very roughly estimate that --- if a critical $\chi_c\!<\!\chi_\infty$ exists --- it should satisfy $(\chi_\infty\!-\!\chi_c)/\chi_\infty\!\lesssim\!0.2$. The finite-size effects observed in our computer experiments are reminiscent of those seen in the fracture of silica pillars~\cite{bonfanti2018damage}, in fuse models~\cite{shekhawat2013damage,driscoll2016role}, and elastic networks~\cite{dussi2020athermal}. Future research efforts should be dedicated towards identifying minimal models that explain these finite-size effects.

\vspace{-0.2cm}



%

\end{document}